\documentclass{webofc}
\usepackage[varg]{txfonts}   
\usepackage{color}
\usepackage{graphicx}
\usepackage{subcaption}
\usepackage{multirow}
\usepackage{lineno}

\newcommand{\bfg }{\begin{figure}[htpb]}
\newcommand{\efg }{\end{figure}}
\newcommand{\bmn }{\begin{minipage}}
\newcommand{\emn }{\end{minipage}}
\newcommand{\bt }{\begin{table}[htpb]}
\newcommand{\et }{\end{table}}

\newcommand{\pA}	{p+Au}
\newcommand{\dA}	{d+Au}

\newcommand{\AuAu}	{Au+Au}

\newcommand{\GeVc}{GeV/$c$ }
\newcommand{\GeVcsq}{GeV/$c^2$ }

\newcommand{ \be }{\begin{equation}}
\newcommand{ \ee }{\end{equation}}
\newcommand{ \bea }{\begin{eqnarray}}
\newcommand{ \eea }{\end{eqnarray}}

\newcommand {\snn}  {\sqrt{s_{_{\rm NN}}}}

\newcommand {\pt}   {p_{T}}

\newcommand {\psiPP}    {\psi_{\rm PP}}

\newcommand {\phires}   {\phi_{\rm res}}

\newcommand {\vres} {v_{2,{\rm res}}}
\newcommand {\Bvec} {\vec{B}}
\newcommand {\psiB}	{\psi_{B}}

\newcommand {\gSS}  {\gamma_{\rm SS}}
\newcommand {\gOS}  {\gamma_{\rm OS}}
\newcommand {\gdel} {\Delta\gamma}
\newcommand {\dg}	{\Delta\gamma}
\newcommand {\dgscale}	{\dg_{\rm scaled}}
\newcommand {\deta}	{\Delta\eta}
\newcommand {\mult}	{N}
\newcommand {\Npv}	{N_{\rm domain}}

\newcommand {\vc}	{v_{2.c}\{2\}}

\newcommand{\rmnum}[1]{\romannumeral #1}
\newcommand{\Rmnum}[1]{\expandafter\@slowromancap\romannumeral #1@}

\newcommand {\mean}[1]  {\langle #1\rangle}

\begin{document}
\title{Chiral magnetic effect search in p+Au, d+Au and Au+Au collisions at RHIC}



\author{\firstname{Jie} \lastname{Zhao}\inst{1}\fnsep\thanks{\email{
			zhao656@purdue.edu
	}} (for the STAR collaboration)
}

\institute{Department of Physics and Astronomy, Purdue University
}

\abstract{%
  Metastable domains of fluctuating topological charges can change the chirality of 
  quarks and induce local parity violation in quantum chromodynamics. This can 
  lead to observable charge separation along the direction of the strong magnetic 
  field produced by spectator protons in relativistic heavy-ion collisions, a phenomenon called the chiral 
  magnetic effect (CME). A major background source for CME measurements using the charge-dependent azimuthal correlator ($\dg$)
  is the intrinsic particle correlations (such as resonance decays) coupled with the 
  azimuthal elliptical anisotropy ($v_{2}$). In heavy-ion collisions, the magnetic field 
  direction and event plane angle are correlated, thus the CME
  and the $v_{2}$-induced background are entangled. 
  In this report, we present two studies from STAR to shed further lights on the background issue. 
  (1) 
  The $\dg$ should be all background in small system p+Au and d+Au collisions,  
  because the event plane angles are dominated by geometry fluctuations uncorrelated to the magnetic field direction. 
  However, significant $\dg$ is observed, comparable to the peripheral Au+Au data, suggesting a background dominance in the latter, 
  and likely also in the mid-central Au+Au collisions where the multiplicity and $v_{2}$ scaled correlator is similar. 
  (2) A new approach is devised to study $\dg$ as a function of the particle pair invariant mass ($m_{inv}$) to identify 
  the resonance backgrounds and hence to extract the possible CME signal. 
  Signal is consistent with zero within uncertainties at high $m_{inv}$.  Signal at low $m_{inv}$,
  extracted from a two-component model assuming smooth mass dependence, is consistent with zero within uncertainties.

}
\maketitle

%
\section{Introduction}
Quark interactions with topological gluon configurations can induce chirality imbalance and local parity violation 
in quantum chromodynamics (QCD)~\cite{Lee:1974ma,Kharzeev:1998kz,Kharzeev:1999cz} .  
In relativistic heavy-ion collisions, this can lead to observable electric charge separation along the direction of the strong magnetic
field produced by spectator protons~\cite{Fukushima:2008xe,Muller:2010jd,Kharzeev:2015znc}. 
This is called the chiral magnetic effect (CME).
An observation of the CME-induced charge separation would confirm a fundamental property of QCD.
The measurements of the charge separation can provide a means
to studying the non-trivial QCD topological structures. 
Extensive theoretical and experimental efforts have been devoted to the search for CME~\cite{Kharzeev:2015znc} . 

The commonly used observable to measure charge separation is the three-point correlator difference~\cite{Voloshin:2004vk}, $\dg\equiv\gOS-\gSS$. 
Here $\gamma=\langle\cos(\alpha+\beta-2\psi_2)\rangle$, $\alpha$ and $\beta$ are the azimuthal angles of two charged particles and $\psi_2$ 
is that of the second-order harmonic plane; $\gOS$ stands for the $\gamma$ of opposite electric charge sign 
(OS) 
and $\gSS$ for that of same-sign pairs 
(SS). 
Significant $\dg$ has indeed been observed in heavy-ion collisions~\cite{Abelev:2009ad,Abelev:2009ac,Adamczyk:2013kcb,Adamczyk:2014mzf,Abelev:2012pa,Ajitanand:2010}. 
One of the difficulties in its interpretation as from the CME is the major charge-dependent background contributions to the $\dg$ observable~\cite{Wang:2009kd,Bzdak:2009fc,Schlichting:2010qia}, 
such as those from resonance decays. The $\dg$ variable is ambiguous between an OS pair from the CME back-to-back 
perpendicular to $\psi_2$ and an OS pair from a resonance decay along $\psi_2$. 
More resonances are produced along the $\psi_2$ than perpendicular to it, 
the relative difference of which is quantified by the elliptical anisotropy parameter $v_2$ of the resonances.
The CME background arises from the coupling of this elliptical anisotropy and the intrinsic decay correlation~\cite{Wang:2016iov}.




The CME and the $v_{2}$-related background are driven by different physics: 
the CME is sensitive to the magnetic field, $\Bvec$, 
while the $v_{2}$-related background is connected to the participant plane, $\psiPP$.
In non-central heavy-ion collisions, the $\psiPP$, although fluctuating~\cite{Alver:2006wh}, is generally 
aligned with the reaction plane (span by the impact parameter direction and the beam), thus generally perpendicular to $\Bvec$. 
The $\dg$ measurement is thus $\emph{entangled}$ by the two contributions: the possible CME and the $v_2$-induced background.
In small-system collisions, however, the $\psiPP$ is determined purely by geometry fluctuations, 
uncorrelated to the impact parameter or the $\Bvec$ direction~\cite{Khachatryan:2016got}. 
As a result any CME would average to zero $\dg$ in small-system collisions. Background sources, on the other hand, 
contribute to small-system collisions similarly as to heavy-ion collisions  
when measured with respect to the event plane reconstructed from mid-rapidity particle momenta (as a proxy to the $\psiPP$). 
Due to the fluctuating nature of the small systems, the event planes reconstructed over a large pseudorapidity gap can be uncorrelated, 
and therefore measurements of mid-rapidity particle correlations with respect to a forward 
event plane are sensitive neither to CME nor to the background.



Recent CMS data show that the correlator signal from p+Pb is comparable to the signal from
Pb+Pb collisions at similar multiplicities~\cite{Khachatryan:2016got,Sirunyan:2017quh}. This indicates significant background contributions in Pb+Pb collisions at LHC energy.
It is predicted that the CME would decrease with the collision energy due to the more rapidly decaying $\Bvec$ at
higher energies~\cite{Kharzeev:2015znc}. Hence, the similarity between small-system and heavy-ion collisions at the LHC may be expected, 
and the situation at RHIC could be different~\cite{Kharzeev:2015znc}. 
In this report, we present results from similar control experiments using p+Au and d+Au collisions at RHIC. 
These results are analyzed as a function of the final-state particle multiplicity 
to shed light on the background to the CME measurements in 
heavy-ion collisions at RHIC. 

Because the main background contributions come from resonance decays, we devise a new analysis approach exploiting the particle 
pair invariant mass, $m_{inv}$, to identify the backgrounds and hence to extract the possible CME signal. 
The $\dg$ signal is reported at large $m_{inv}$ where resonance background contributions are small. 
A two-component model fit to the low mass region is also reported along with the fitted possible CME signals. 


\section{Experiment setup and data analysis}
The data reported here were collected by the STAR
experiment~\cite{Ackermann:2002ad} at Brookhaven National Laboratory from 2003 (d+Au), 2011 (Au+Au) and 2015 (p+Au).
The main subsystems used for the data analysis 
are the Time Projection Chamber (TPC)~\cite{Anderson:2003ur} and the Time-Of-Flight detector
(TOF)~\cite{Llope:2003ti}, both with 2$\pi$ azimuthal coverage at mid-rapidity. 

Charged particle identification is performed by ionization energy
loss, $dE/dx$, measured by TPC and/or particle velocity measured by TOF.
The correlations are reported for charged particles with pseudorapidity $|\eta| < 1$ and transverse
momentum $0.2< p_{T} <2.0$ \GeVc, and for identified $\pi^{\pm}$ with $0.2< p_{T} <1.8$ \GeVc.

We use the three-particle correlator: 
\begin{equation}
	\gamma \equiv \langle \cos(\phi_{\alpha}+\phi_{\beta}-2\psi_{2})\rangle = \langle \cos(\phi_{\alpha} + \phi_{\beta} - 2\phi_{c})\rangle/v_{2}. 
	\label{eqThreeCtor0}
\end{equation}
Here $\phi_{\alpha}$ and $\phi_{\beta}$ are the azimuthal angle of the two particles, $\phi_{c}$ is that of the third particle 
which serves as a measure of the $\psiPP$. The imprecision in 
determining the $\psiPP$ by a single particle is corrected by
the resolution factor, equal to the particle's elliptic 
flow anisotropy $v_{2,c}$. 
The $\dg$ from the difference of $\gOS$ and $\gSS$ correlator is used to quantify the charge dependent signal.  
We use particles in the full TPC 
acceptance ($|\eta| < 1$) for all three particles. 
No $\eta$ gap ($\Delta\eta$) is applied between any pair among the three particles. 
The $v_{2,c}$ used for resolution correction should, ideally, 
be free of non-flow. We obtain $v_{2,c}$ by the two-particle cumulant, 
applying a cut of $\Delta\eta > 1.0$ to reduce non-flow contaminations.

\section{Results from small systems}
\begin{figure}[htbp!]
	\centering 
	\includegraphics[width=7.0cm]{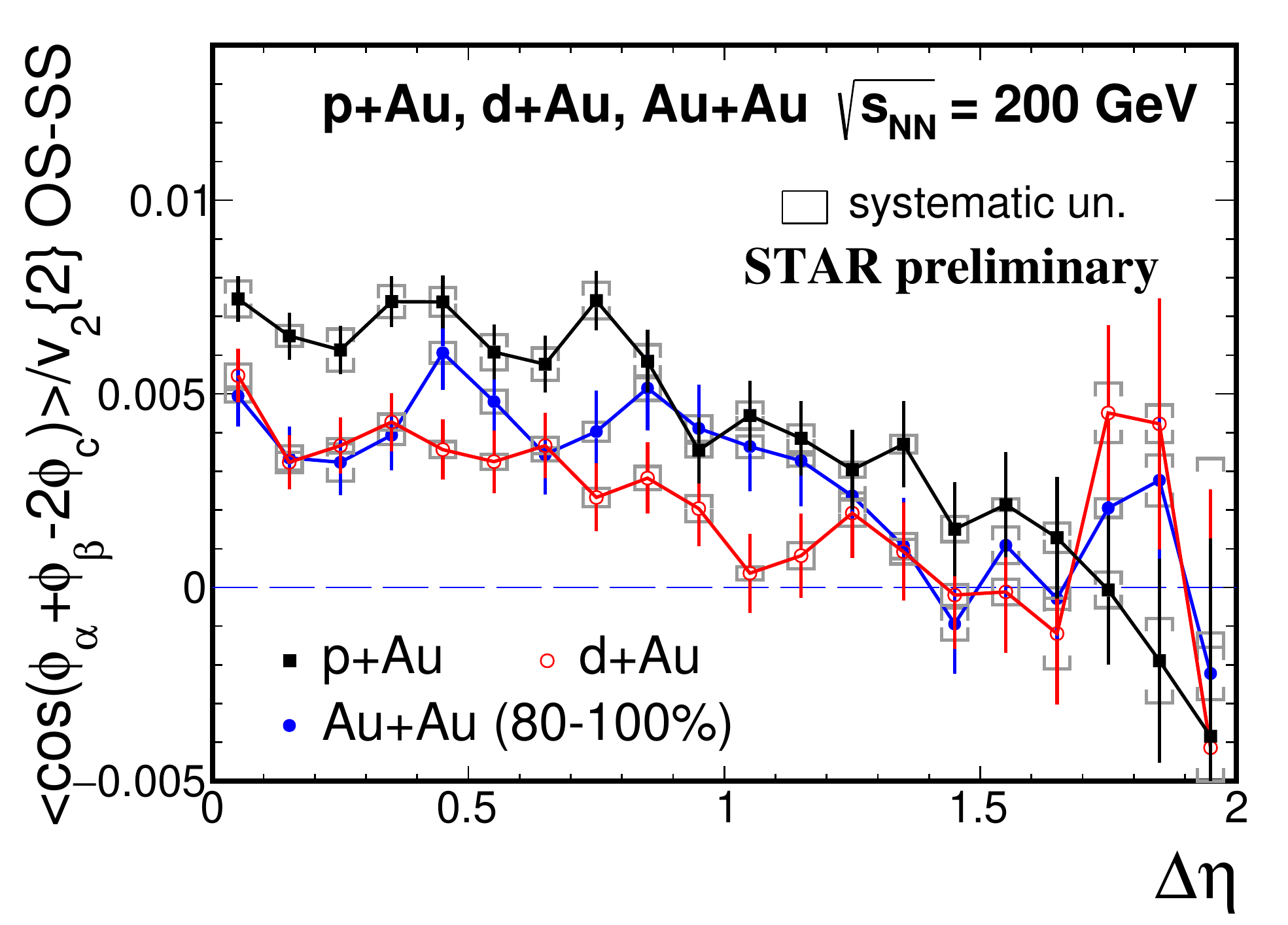} 
	\caption{(Color online)
		The $\gdel$ in p+Au and d+Au collisions as a function of $\Delta\eta$ between particle $\alpha$ and $\beta$, compared to that in peripheral Au+Au collisions.  
	}   
	\label{fig3}
\end{figure}

The three-particle correlator from small systems are analyzed as functions of rapidity-gap and multiplicity to shed light on the 
background contaminations to the CME measurements in heavy-ion collisions.  
Figure~\ref{fig3} shows $\dg$ as a function of the $\Delta\eta$ between particle $\alpha$ and $\beta$ in \pA\ and \dA\ collisions, 
and in peripheral \AuAu\ collisions for comparison. The distributions from \pA\ and \dA\ collisions are similar to those in peripheral \AuAu\ collisions.

\begin{figure}[htbp!]
	\centering 
	\includegraphics[width=7.0cm]{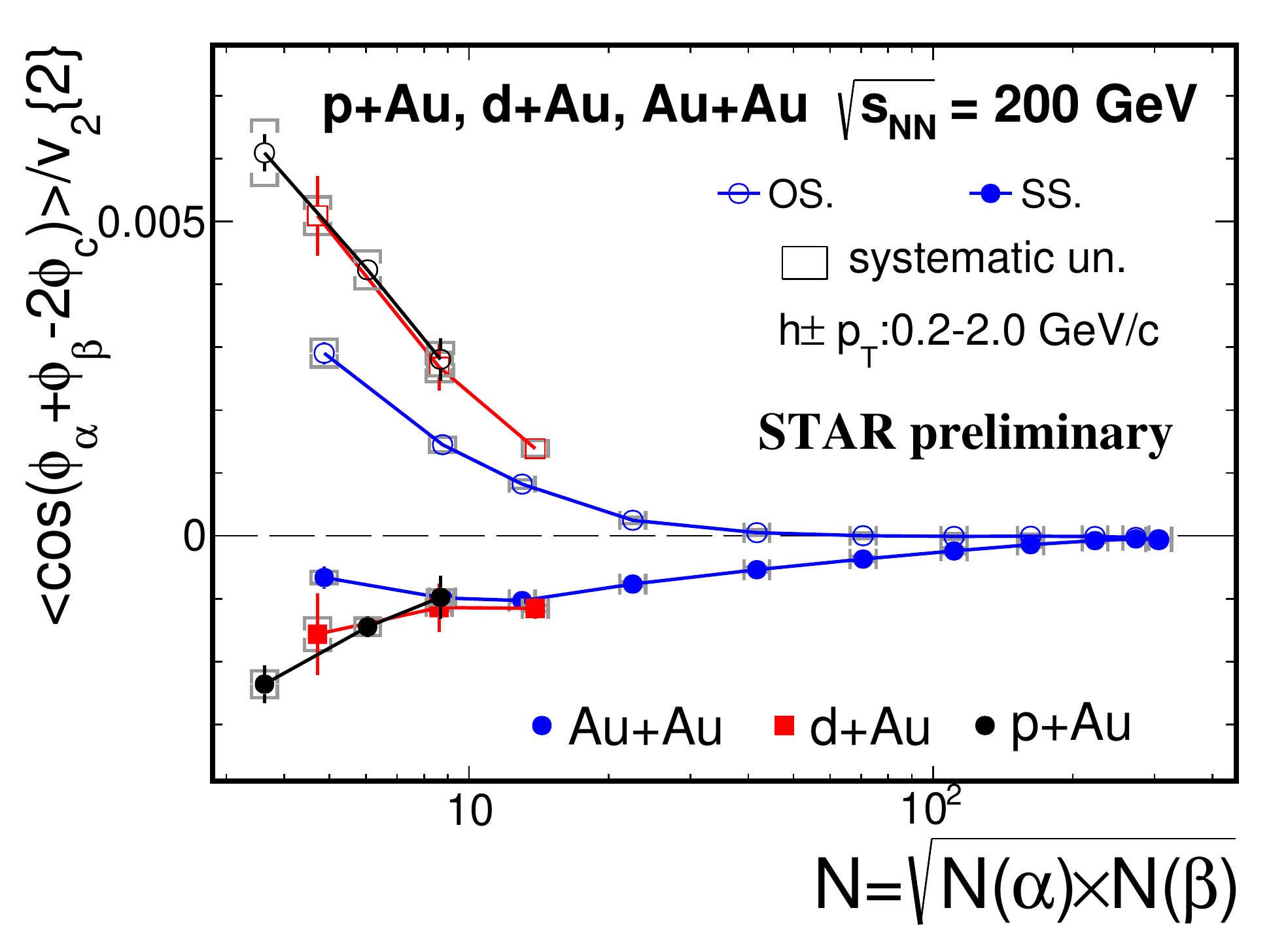} 
	\includegraphics[width=7.0cm]{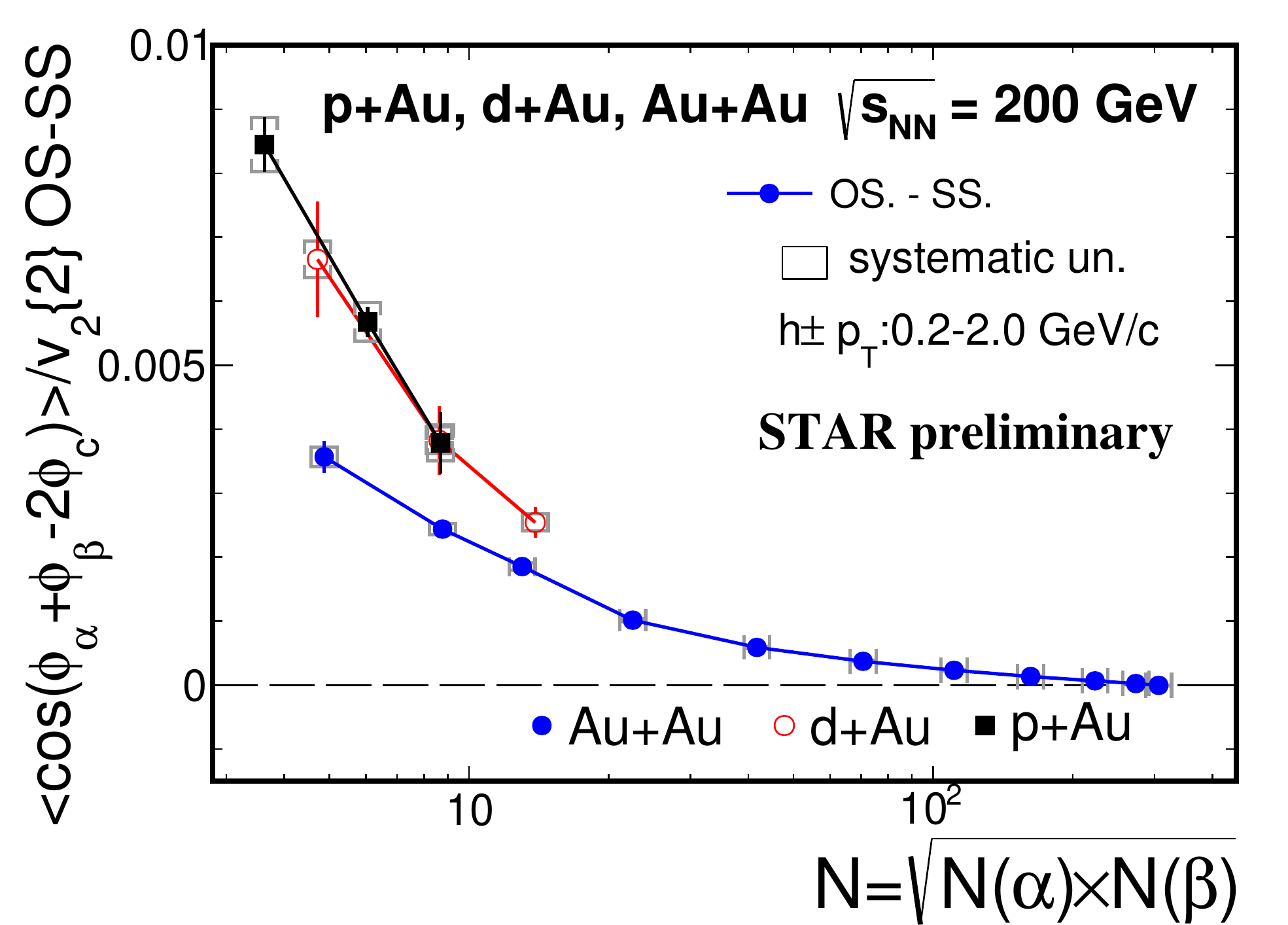}
	\caption{(Color online)
		The $\gSS$, $\gOS$ (Left panel) and $\gdel$ (Right panel) correlators in p+Au and d+Au collisions as a function of multiplicity, compared to those in
		Au+Au collisions. Particles $\alpha$, $\beta$ and $c$ 
		are from the TPC pseudorapidity coverage of $|\eta|<1$ with no $\eta$ gap applied.
		The $v_{2,c}\{2\}$ is obtained by two-particle cumulant with $\eta$ gap of $\Delta\eta > 1.0$. Statistical uncertainties 
		are shown by the vertical bars and systematic uncertainties are shown by the caps.
	}   
	\label{fig1}
\end{figure}

Figure~\ref{fig1} (Left) shows the $\gSS$ and $\gOS$ results as functions of particle multiplicity, $\mult$, in \pA\ and \dA\ collisions at $\snn=200$~GeV. 
Here $\mult$ is taken as the geometric mean of the multiplicities of particle $\alpha$ and $\beta$.
The systematic uncertainties on $\gamma$ are estimated by varying the track quality cuts, the correction method used for the detector non-uniform azimuthal acceptance,
and the $\pt$ range of the reference particle $c$. 
For comparison the corresponding \AuAu\ results are also shown. 
The trends of the correlator magnitudes are similar, decreasing with increasing $\mult$. The $\gSS$ results seem to follow a smooth trend in $\mult$ over all systems. 
The $\gOS$ results are less so; the small system data appear to differ somewhat from the heavy-ion data over the range in which they overlap in $\mult$. 
Since the particle pseudorapidity density, $dN/d\eta$, is asymmetric in small systems, there are relatively more small $\deta$ pairs than in \AuAu\ collisions, 
hence larger correlations. We have checked this effect by weighting particles in computing the correlator such that the weighted $dN/d\eta$ is symmetric, 
and found it insufficient to account for the observed difference.

\begin{figure}[htbp!]
	\centering 
	\includegraphics[width=7.0cm]{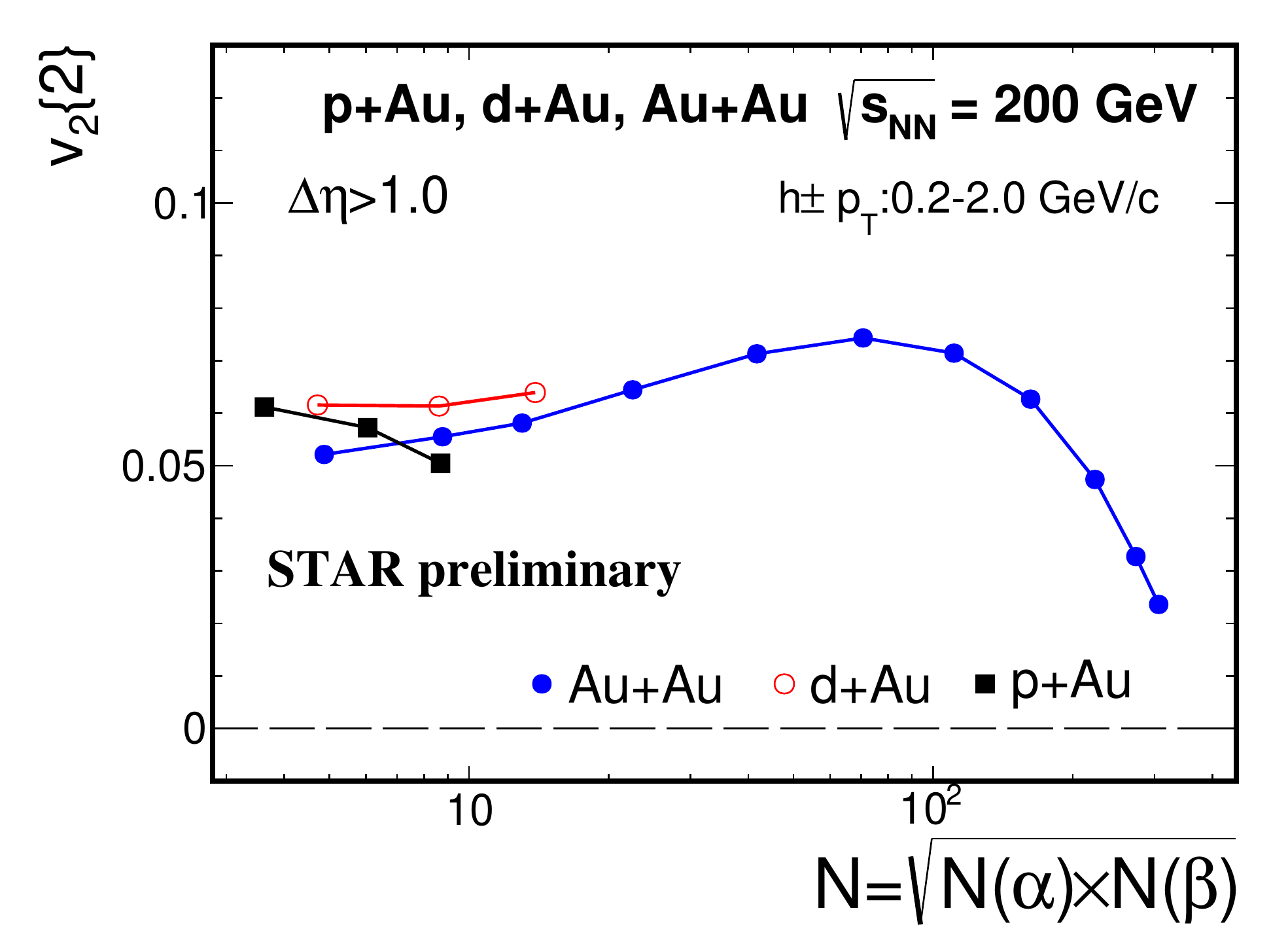} 
	\includegraphics[width=7.0cm]{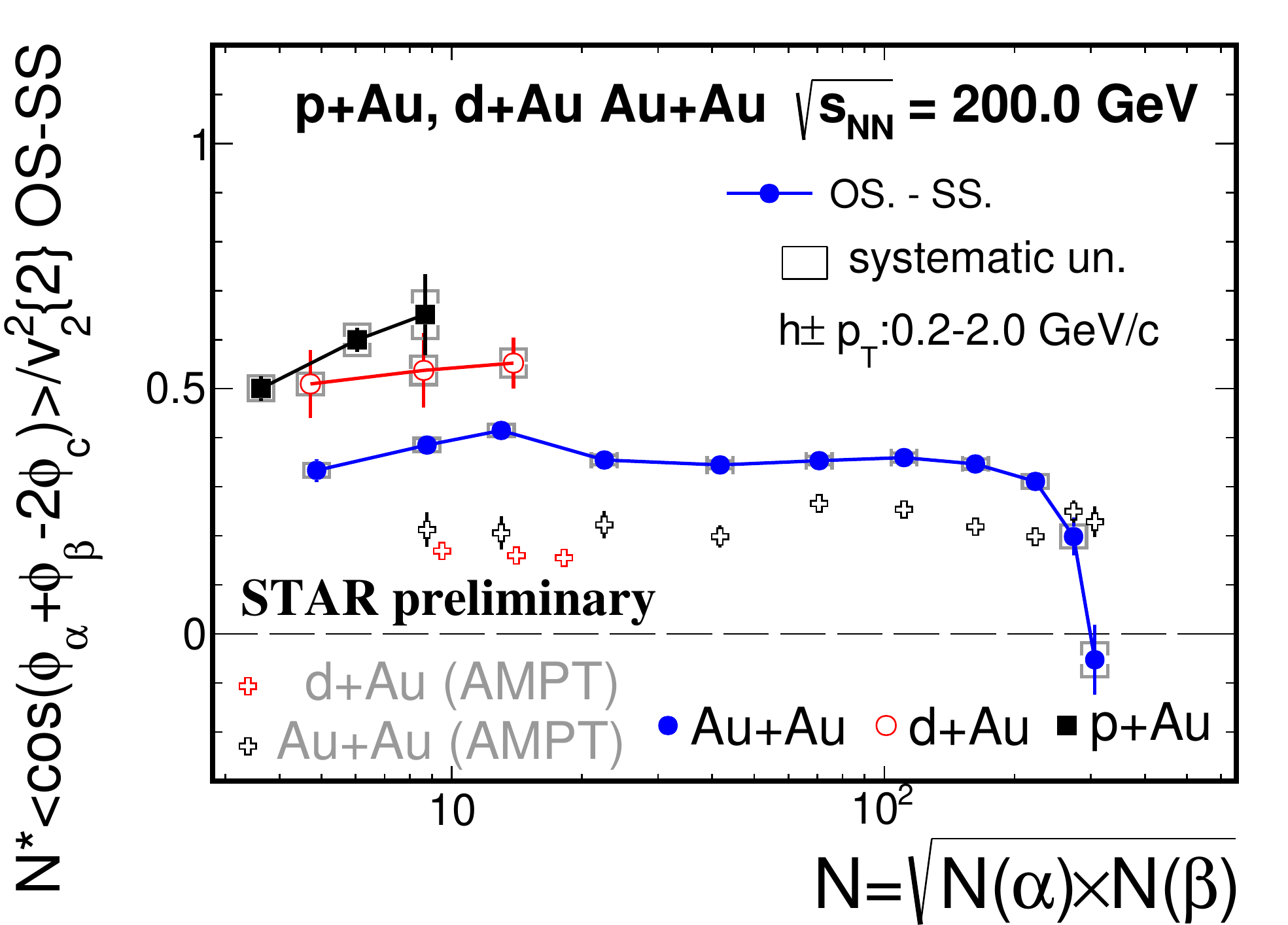}
	\caption{(Color online)
		(Left panel) The measured two-particle cumulant $v_{2,c}\{2\}$ of charged particles ($0.2<\pt<2.0$ \GeVc) with $\eta$ gap of 1.0, 
		and (Right panel) the scaled three-particle correlator difference in p+Au and d+Au collisions as a function of $\mult$, 
		compared to those in Au+Au collisions. AMPT simulation results for d+Au and Au+Au are also plotted for comparison.
	}   
	\label{fig2}
\end{figure}

Similar to LHC, the small system data at RHIC are found to be comparable to Au+Au results at similar multiplicities.
Since the \pA\ and \dA\ data are all backgrounds, the $\dg$ should be approximately proportional to the averaged $v_2$ of the background sources, 
in turn the $v_2$ of final-state particles. It should also be proportional to the number of background sources, 
and, because $\dg$ is a pair-wise average, inversely proportional to the total number of pairs. 
As the number of background sources likely scales with multiplicity, we have $\dg\propto v_2/\mult$.
Therefore, to gain more insight, we scale the $\dg$ by $\mult/v_2$:
\begin{equation}
	\dgscale=\dg\times\mult/v_2\,.
	\label{eq:scale}
\end{equation}
Since there is no distinction between the two particles ($\alpha, \beta$) and particle $c$, the $v_2$ in Eq.~(\ref{eq:scale}) is the same as $\vc$.
The $v_{2}\{2\}$ with a $\eta$ gap of 1.0 is shown as a function of $\mult$ in \pA, \dA, and \AuAu\ collisions in Fig.~\ref{fig2} (Left).
Figure~\ref{fig2} (Right) shows the scaled $\dgscale$ as a function of $\mult$ in \pA\ and \dA\ collisions, and compares to that in \AuAu\ collisions. 
AMPT simulation results for d+Au and Au+Au are also plotted for comparison. The AMPT simulations can account for about $2/3$ of the STAR data, 
and are approximately constant over $\mult$. 
The $\dgscale$ in \pA\ and \dA\ collisions are compatible or even larger than that in \AuAu\ collisions. Since in \pA\ and \dA\ collisions only the background is present, the data suggest that the peripheral \AuAu\ measurement may be largely, if not all, background.
For both small-system and heavy-ion collisions, the $\dgscale$ is approximately constant over $\mult$. 
It may not be strictly constant because
the correlations caused by decays ($\dg_{\rm bkgd}\propto\mean{\cos(\alpha+\beta-2\phires)}\times\vres$), 
depends on the $\mean{\cos(\alpha+\beta-2\phires)}$ which is determined by the parent kinematics and can be somewhat $\mult$-dependent. 
Given that the background is large suggested by the \pA\ and \dA\ data, the approximate $\mult$-independent $\dgscale$ in \AuAu\ collisions is consistent with the background scenario.

In terms of CME, on the other hand, the number of local topological domains, $\Npv$, is likely proportional to the collision volume, so $\Npv\propto\mult$. 
Since the topological charge randomly fluctuates, the charge asymmetry $a_1\propto\sqrt{\Npv}/N$. 
As a result, the $\dg\propto a_1^2$ would be inversely proportional to $\mult$, similar to the background scenario. 
The CME-induced charge separation magnitude is expected to be proportional to $\mean{B^2\cos2(\psiB-\psi_2)}$~\cite{Kharzeev:2015znc}, 
where $\psiB$ is the azimuthal angle of $\Bvec$. The quantity, $\mean{B^2\cos2(\psiB-\psi_2)}$, may have similar $\mult$-dependence as the $v_2$ 
but the present theoretical uncertainties are large~\cite{Deng:2012pc,Bloczynski:2012en}. 
Thus, the $\mult$-dependence of the CME and the background may, unfortunately, be similar. 
Given the present uncertainties, a small finite CME signal is not disallowed in the measured $\dg$ in heavy-ion collisions.
The present analysis does not currently allow conclusive statements to be made regarding the presence of CME.

\section{Results from the invariant mass method}

\begin{figure}[htpb]
	\centering
	\begin{subfigure}[p]{0.43\textwidth}
		\includegraphics[width=\textwidth]{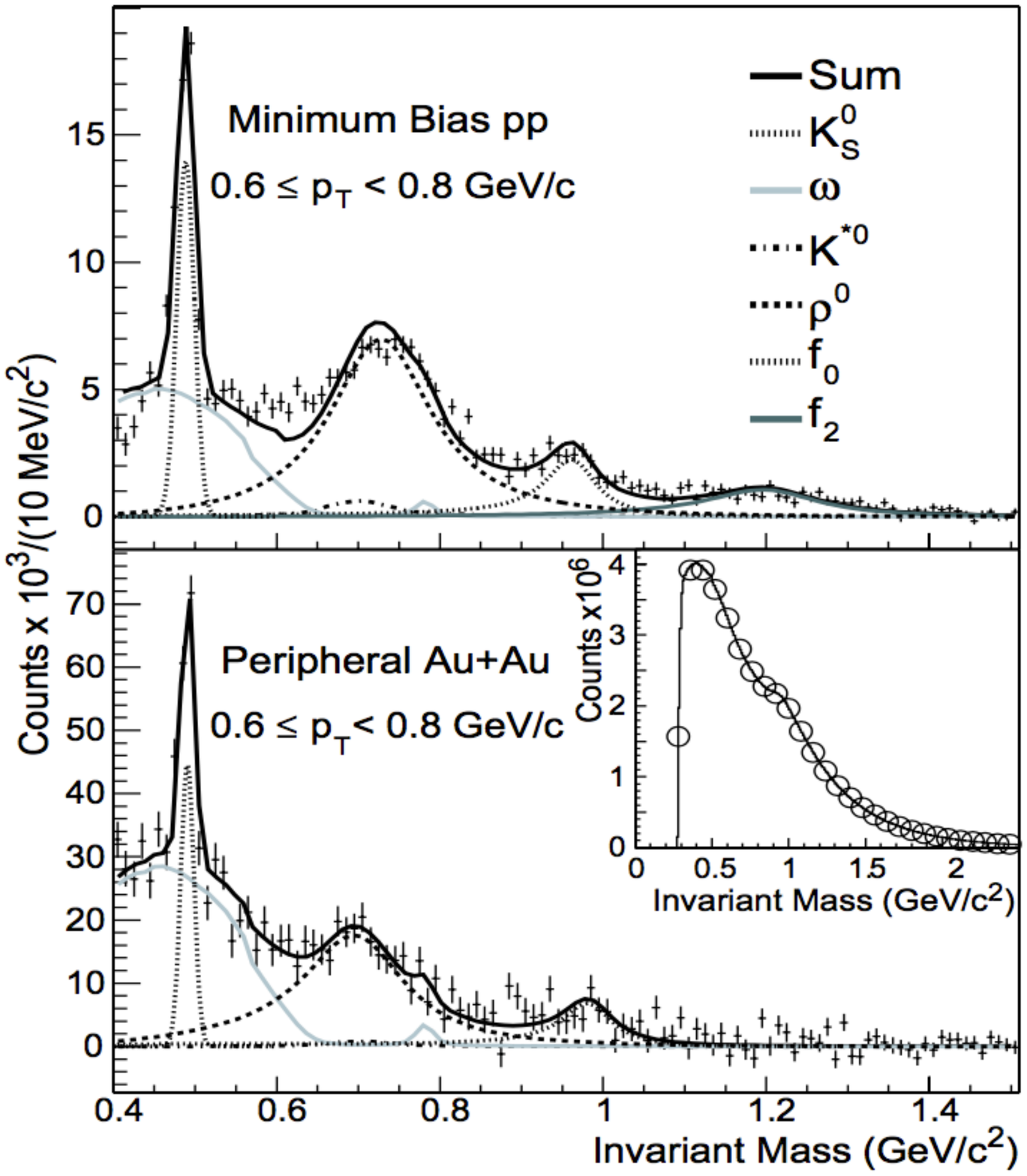}
	\end{subfigure}
	\begin{subfigure}[p]{0.56\textwidth}
		\includegraphics[width=\textwidth]{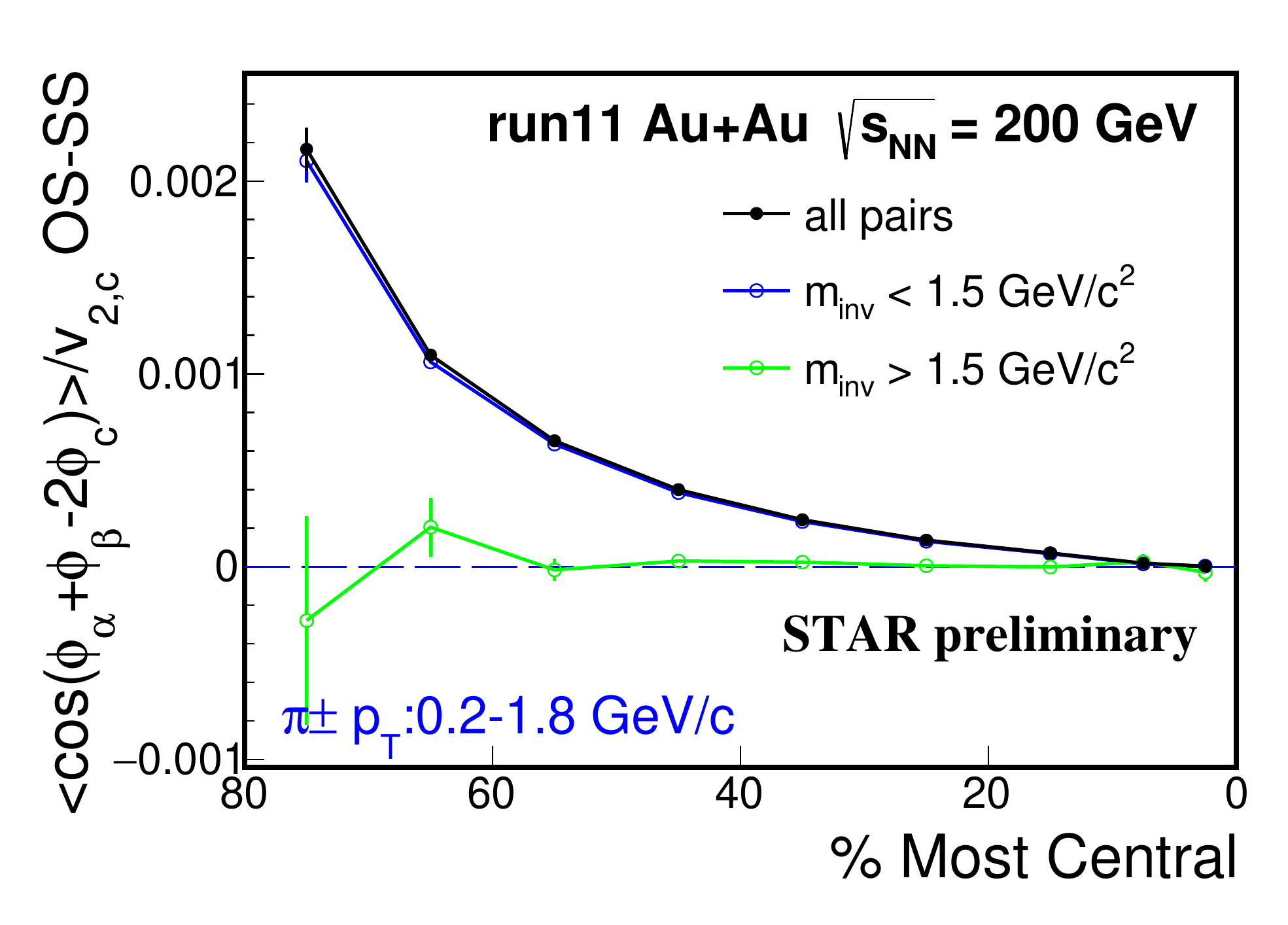}
	\end{subfigure}
	\caption{
		(Left panel) The raw $\pi$-$\pi$ invariant mass ($m_{inv}$) distributions after subtraction of the like-sign reference distribution for minimum bias p+p (top) 
		and peripheral Au+Au (bottom) collisions. 
		The insert plot corresponds to the raw $\pi^{+}$-$\pi^{-}$ $m_{inv}$ (solid line) and the like-sign reference 
		distributions (open circles) for peripheral Au+Au collisions~\cite{Adams:2003cc}.
		(Right panel) The inclusive $\gdel$ over all mass (black) and at $m_{inv} > 1.5$ \GeVcsq (green) as a function of centrality in Au+Au collisions at 200 GeV. 
	}
	\label{fig4}
\end{figure}

As discussed in the introduction, the major background arises from resonance decay correlations coupled with the elliptical anisotropy. 
Figure~\ref{fig4} (Left) shows the OS and SS $\pi\pi$ pair difference as a function of $m_{inv}$ 
from minimum bias p+p and peripheral Au+Au collisions at 200 GeV~\cite{Adams:2003cc}. 
Many resonances have broad mass distributions~\cite{Agashe:2014kda}. 
They are often identified statistically in relativistic heavy ion collisions.
However statistical identification of resonances does not help eliminate their contribution to the $\dg$ variable. 
Most of the $\pi\pi$ resonances are located in the low $m_{inv}$ region~\cite{Adams:2003cc}. 
With increasing mass, the data show that resonance contributions to 
the difference between OS and SS pairs decrease.
It is possible to exclude them ``entirely'' by applying a lower cut on $m_{inv}$. 
In the low mass region we analyze the $\dg$ correlator as a function of $m_{inv}$, 
which may help to understand the background sources and isolate the possible CME~\cite{Zhao:2017nfq}. 


Figure~\ref{fig4} (Right) shows the $\dg$ results with and without applying an invariant mass cut of $m_{inv}>1.5$~\GeVcsq. 
The results are summarized in Table~\ref{table1}. 
With the $m_{inv}$ cut, the $\dg$ is significantly reduced from the inclusive measurement. 
The $\dg$ with the large $m_{inv}$ cut is consistent with zero within the current uncertainty.

\bt
\caption{The inclusive $\dg$ over all mass and $\dg$ at $m_{inv}>1.5$ \GeVcsq for different centralities in Au+Au collisions at 200 GeV.} 
\centering
\begin{tabular}{lccc}
\hline
\hline
Centrality  &  $\dg$ in all mass (A) &  $\dg$ at $m_{inv}>1.5$ \GeVcsq (B) & B/A \\ 
\hline
50-80\% & $(7.45\pm0.21)\times10^{-4}$   & $(1.3\pm5.7)\times10^{-5}$ & $(1.8\pm7.6)\%$ \\
20-50\% & $(1.82\pm0.03)\times10^{-4}$   & $(7.7\pm9.0)\times10^{-6}$ & $(4.3\pm4.9)\%$ \\
 0-20\% & $(3.70\pm0.67)\times10^{-5}$   & $(-0.1\pm1.8)\times10^{-5}$ & $(-3.8\pm49)\%$ \\
\hline
\hline
\end{tabular}
\label{table1}
\et

\begin{figure}[htbp!]
	\centering 
	\includegraphics[width=7.0cm]{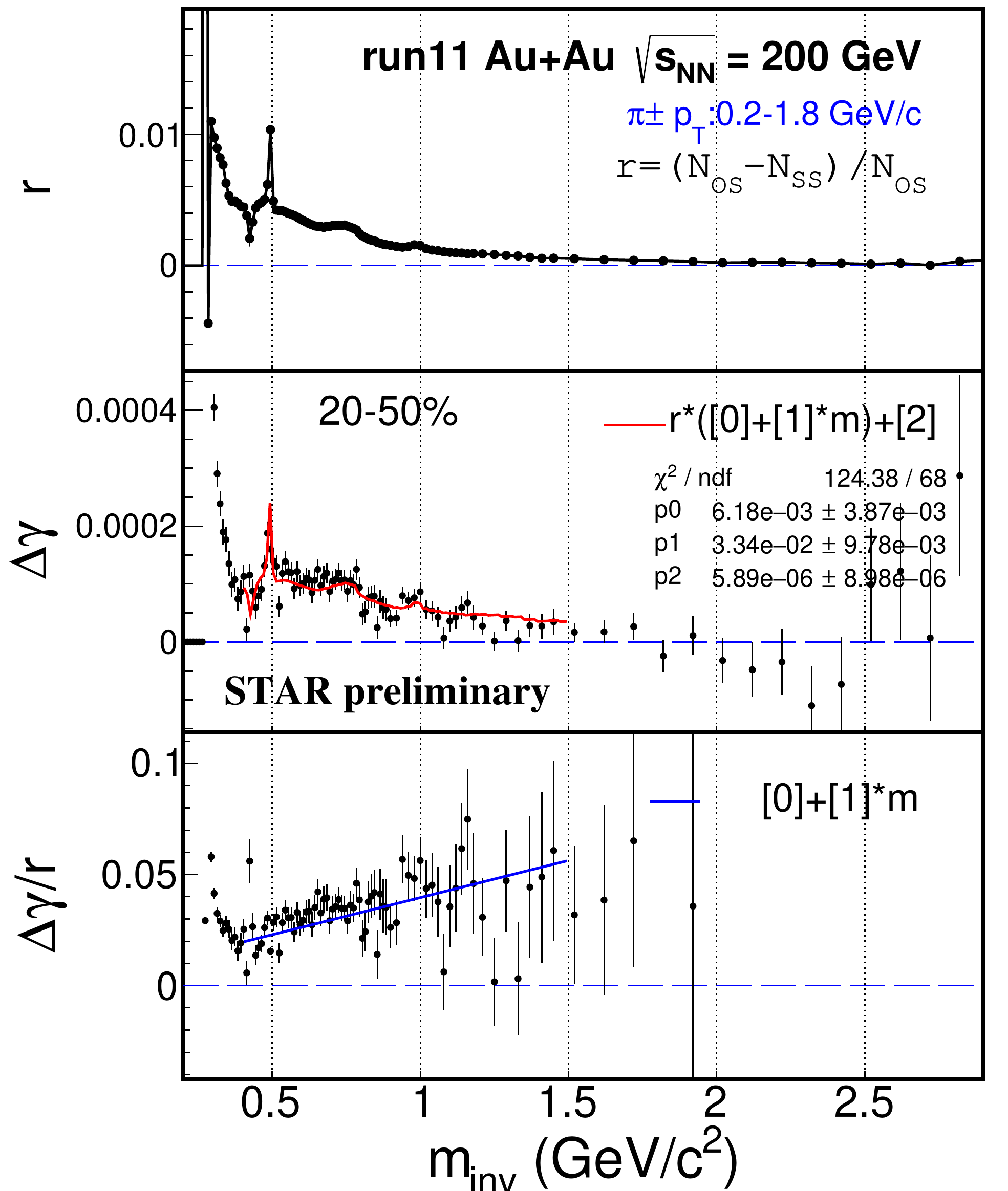} 
	\caption{(Color online)
		Pair invariant mass ($m_{inv}$) dependence of the relative excess of OS over SS charged $\pi$ pair multiplicity, $r=(N_{OS}-N_{SS})/N_{OS}$ (top panel), 
		event-plane dependent azimuthal correlator difference, $\gdel=\gOS-\gSS$ (middle panel), and the ratio of $\gdel/r$ (bottom panel) in 20-50$\%$ 
		Au+Au collisions at 200 GeV. Errors shown are statistical.
		The red curve in the middle panel shows the two-component model fit assuming a constant CME contribution independent of $m_{inv}$;
		The blue curve in the bottom panel shows the corresponding resonance response function.
	}   
	\label{fig5}
\end{figure}

CME is expected to be a low $\pt$ phenomenon~\cite{Kharzeev:2007jp,Abelev:2009ad}; its contribution to high mass may be small. 
In order to extract CME at low mass, resonance contributions need to be subtracted.
To this end, we show in Fig.~\ref{fig5} (a) the relative OS and SS pair abundance difference, $r=(N_{OS}-N_{SS})/N_{OS}$, 
and in Fig.~\ref{fig5} (b) the $\dg$ correlator as a function of $m_{inv}$  
from mid-central (20-50\%) Au+Au collisions at 200~GeV.
The data show resonance structures in $r$ and $\dg$ as functions of $m_{inv}$: 
a clear resonance peak from $K_{s}^{0}$ decay is observed, and possible $\rho$ and $f^{0}$ peaks are also visible.
The $\dg$ correlator traces the distributions of those resonances. 

The $\dg$ in Fig.~\ref{fig5} (b) may be composed of two components, a resonance decay background and a CME signal: 
\begin{equation}
	\dg=r(m_{inv})\cdot R(m_{inv}) + \rm{CME}(\it{m_{inv}}).
	\label{eqFit}
\end{equation}
Here CME($m_{inv}$) represents the CME contribution. 
The background depends on $r(m_{inv})$, with a smooth response function $R(m_{inv})$, so the background component is peaky in $m_{inv}$. 
We assume that the CME component is smooth in $m_{inv}$. 
If the CME contribution was appreciable, then the ratio of $\dg/r$ shown in Fig.~\ref{fig5} (c) would reveal a structure resembling the inverse shape of $r$~\cite{Zhao:2017nfq}. 
However, a more or less smooth dependence is observed;
no evidence of an inverse shape of the resonance mass distribution is observed in the ratio of $\dg/r$, suggesting insignificant CME signal contributions.

In order to isolate the possible CME from the resonance contributions, 
the two-component model is used to fit the $\dg$ as a function of $m_{inv}$. 
We use the first-order polynomial function for $R(m_{inv})$,
motivated by the data in Fig.~\ref{fig5} (c) and MC simulation~\cite{Zhao:2017nfq}.
At present, no theoretical calculation is available on the $m_{inv}$ dependence of the CME, 
therefore we consider two cases: (\rmnum{1}) a constant CME distribution independent of $m_{inv}$, and (\rmnum{2}) an exponential CME distribution in $m_{inv}$.
The curve superimposed in Fig.~\ref{fig5} (b) shows the fit of case (\rmnum{1}).
The extracted average $\dg$ from CME contributions, and their strengths relative to the inclusive $\dg$ measurement are tabulated in Table~\ref{table2}. 
The extracted CME signals from the two-component model fits are at present consistent with zero within uncertainties.
Future theoretical calculations of the CME mass dependence would be valuable.

\bt
\caption{
	Average $\dg$ signal (corresponding to the CME contribution in the two-component fit model) 
	extracted from the model fit
	at $m_{inv} < 1.5$ \GeVcsq 
	in mid-central ($20-50\%$) Au+Au collisions at 200 GeV, 
	with two assumptions for the CME $m_{inv}$ dependence: a constant independent of $m_{inv}$ and an exponential in $m_{inv}$.
} 
\centering
\begin{tabular}{lcc}
\hline
\hline
$\dg$ (inclusive)          &   \multicolumn{2}{c}{$(1.82\pm0.03)\times10^{-4}$}  \\ 
\hline
& constant CME             & exponential CME in $m_{inv}$  \\
average signal $\dg$ (fit) & $(5.9\pm9.0)\times10^{-6}$   & $(3.0\pm2.0)\times10^{-5}$ \\
fit/inclusive              & $(3.2\pm4.9)\%$              & $(16\pm11)\%$ \\
\hline
\hline
\end{tabular}
\label{table2}
\et

\section{Summary}
The chiral magnetic effect (CME) can produce charge separation perpendicular to the reaction plane. 
Charge separation measurements by the three-particle correlators ($\dg$) are contaminated by a major background arising from particle 
correlations coupled with elliptical anisotropy ($v_2$). 
To shed more light on the background and to reduce/eliminate background contamination in charge separation measurements, 
we have studied the small-system p+Au and d+Au collisions in comparison to Au+Au collisions, 
and the $\dg$ correlator as a function of the particle pair invariant mass ($m_{inv}$).

With respect to the second-order harmonic plane ($\psi_{2}$), the p+Au and d+Au charge dependent 
correlations are backgrounds. Peripheral Au+Au data are similar
to those of p+Au and d+Au. The scaled correlator ($\dg\times\mult/v_2$) from peripheral to mid-central Au+Au
collisions is approximately constant over multiplicity. 
Similar dependence is found in AMPT.
These data indicate a dominant contribution from background to the
peripheral and mid-central heavy-ion collisions, and do not show clear evidence for the
presence of the CME in those collisions.

A new method exploiting the particle pair invariant mass is used to identify the resonance background and the possible CME.
In order to exclude the resonance contributions, we apply a lower cut on the $m_{inv}$. 
At high mass ($m_{inv}>1.5$ \GeVcsq), $\dg$ is consistent with zero within uncertainty.
In the low mass region ($m_{inv}<1.5$ \GeVcsq), data show resonance structures in $\dg$ as function of $m_{inv}$.
A two-component fit is devised where the background component is peaky following the resonance contributions 
and we assume that the CME signal is smooth in $m_{inv}$. Two functional forms are assumed in the present study for the CME as function of $m_{inv}$, 
a constant and an exponential. 
The extracted CME signals from the two-component model fit are consistent with zero within the current uncertainties.
Theoretical guidance on the mass dependence of CME would be valuable to further our understanding.

\section{Acknowledgments}
This work was partly supported by the U.S. Department of Energy (Grant No. de-sc0012910).


\bibliographystyle{unsrt}
\bibliography{./ref}
\end{document}